# Raman and Infrared spectra of $(BaF_2)_n$ (n=1-6) clusters

Ratnesh K. Pandey[1, 2*], Kevin Waters[2], Sandeep Nigam[3], Ravindra Pandey[2], Avinash C. Pandey[1]

[1]*Nanotechnology Application Centre, University of Allahabad, Allahabad, India, 211002.*
[2]*Michigan Technological University, Houghton, Michigan, USA, 49931.*
[3]*Chemistry Division, Bhabha Atomic Research Centre, Trombay, Mumbai, India, 400085.*

[*]*pandeyratneshk@gmail.com*

**Abstract.** The vibrational properties of alkaline-earth metal fluoride clusters $(BaF_2)_n$ (n=1-6) are investigated in the framework of density functional theory. The calculated Raman and Infrared (IR) spectra reveals shift in Raman and IR peak position towards lower frequency region with the increase in the cluster size. Further the calculated spectra have been compared with the experimental vibrational spectra of bulk $BaF_2$ crystal. Even though the smaller size cluster lacks translational symmetry, the structural and vibrational characteristic of $(BaF_2)_{5-6}$ are nearer to bulk counterpart.

## INTRODUCTION

In recent times, significant efforts have been made to study the size-dependent evolutionary properties of small clusters and, in particular, how their properties converge to corresponding bulk values. Specifically, ionic clusters, such as alkali halide, alkaline-earth oxides or alkaline-earth halides are likely to have the bonding characteristics which remain similar throughout all sizes implying that the stoichiometric ionic clusters may have stable bulk-like configurations even at nanoscale.

Barium fluoride has a fluorite cubic lattice structure in which a cube of $Ba^{2+}$ ions is surrounded by a cube of $F^-$ ions. Due to its large band gap, $BaF_2$ is commonly used as the choice material for the transmission window in the optical instruments. $BaF_2$ has been extensively characterized in its bulk form[1, 2], though no such characterization is reported at the nanoscale. Recently, the cluster emission was observed after swift heavy ion bombardment on nanometric thin films of $BaF_2$[3] which led us to initiate a detailed investigation of the physical and chemical properties of the clusters of $BaF_2$. In the present article we focus on Raman and infrared (IR) spectrum of $BaF_2$ clusters to investigate their size dependency. It is further attempted to compare their behavior with the bulk counterpart.

## COMPUTATIONAL DETAILS

First principles calculations were performed on $BaF_2$ clusters in the framework of density functional theory using Gaussian 09 program[4]. The gradient-corrected B3LYP functional form (i.e. Becke's 3-parameter hybrid exchange functional and Lee, Yang, and Parr correlation functional) was employed. The Los Alamos National Laboratory (LanL2DZ) basis sets for Ba and the 6-31G* basis set for F was used in these calculations. All of the clusters being analyzed in this study were fully optimized and the energy and density convergence criteria were fixed at $10^{-6}$ Ha and $10^{-6}$ e /Bohr$^3$, respectively. The vibrational frequencies under the harmonic approximation, with analytic force constant were calculated.

## RESULTS AND DISCUSSION

Several linear, planar and non-planar structures of $(BaF_2)_n$ (n=1-6) clusters are considered for the geometry optimizations. Fig.1 shows the ground state configurations of $(BaF_2)_n$ clusters which were used for frequency calculations. Various vibrational frequencies of $(BaF_2)_n$ (n=1-6) clusters are presented in figure-2. For n=1, only few modes are present there, however as the number of BaF$_2$ unit increases, the number of vibrational modes also increases.

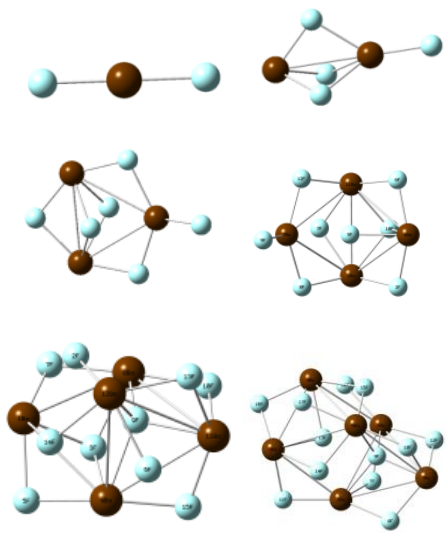

**FIG 1.** The ground state of (BaF$_2$)$_n$ (n=1-6) clusters

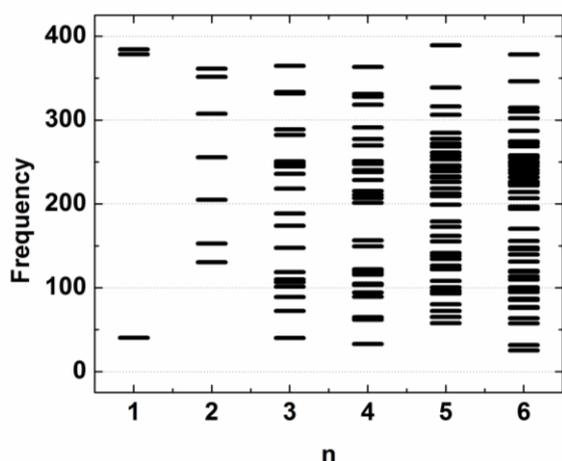

**FIG 2.** Vibrational frequencies of (BaF$_2$)$_n$ (n=1-6) clusters

Fig.3 shows the calculated Raman spectra of (BaF$_2$)$_n$ (n=1-6) clusters. For n=1, there is only one peak occurring at 378 cm$^{-1}$. As the number of BaF$_2$ unit increases, the peaks shift towards lower frequencies. This has been attributed due to the fact that the Ba-F bond length increases systematically with increase in the cluster size (From 2.33 to 2.55A). Since bond elongation in turn weakens the bond hence the peak position shifts to lower energy region. The same analogy can be applied to infrared (IR) spectra shown in Fig. 4 which shows almost the similar trend as Raman spectra.

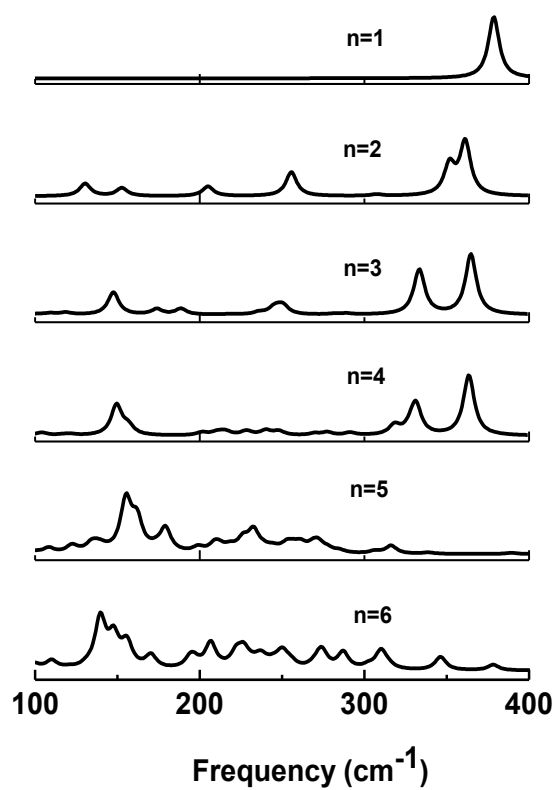

**FIG 3.** Raman spectra of (BaF$_2$)$_n$ (n=1-6) clusters

Both Raman and IR spectra look similar for n=1-3. As the cluster size increases, more number of peaks appears in IR spectra relative to that in Raman spectra. Since fluorine atom has lower mass than the Ba atom, therefore it is expected that F atom will be contributing relatively more in vibration modes appearing in the studied energy regime of the spectra. As the number of F atoms is increasing with the cluster size, accordingly the number of peaks in the vibrational spectra is also increasing. Since co-ordination number of individual atom plays important role in rationalizing the bond-distance and vibrational modes, we have analyzed the co-ordination number of F atom with respect to cluster size. We have found that for n = 2-4, coordination numbers are 2 and 1 for fluorine atoms, but for n=5 and 6, maximum coordination number value becomes 4, which is co-ordination number of F atom in bulk BaF$_2$. Further it is observed that symmetry of individual cluster decreases on increase in cluster size. For example (BaF$_2$)$_2$ cluster has C$_{3v}$ point group symmetry, while (BaF$_2$)$_6$ cluster has C$_s$ point group symmetry. Increase in the co-ordination number and reduction in symmetry, results in complicated intensity pattern in IR and vibrational spectra.

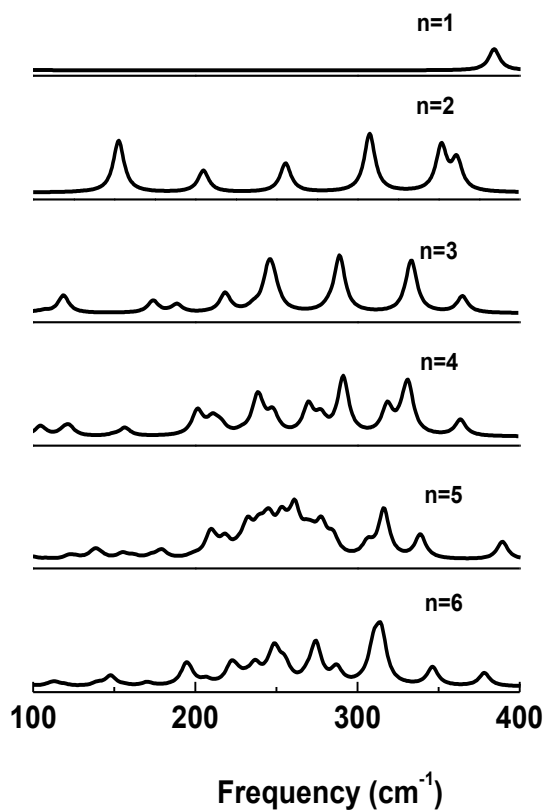

**FIG 4.** Infrared spectra of $(BaF_2)_n$ (n=1-6) clusters.

It is worth to mention here that, in bulk $BaF_2$ crystal, F atom positioned itself in tetrahedral hole, with Ba-F distance as 2.66 Å[5]. The reported[6] central zone frequency for IR active mode is 190-200 cm$^{-1}$ and for Raman active mode, it is 240-260 cm$^{-1}$. It is further reported[7] that, doping of pure $BaF_2$ cluster, leads to appearance of many non-center zone modes in the IR and Raman spectra. These additional modes are activated by loss of translational symmetry and observed at nearby frequency regions. In the present case, the $(BaF_2)_{5-6}$ cluster not only have the F atom`s average coordination number as four ( equal to the bulk value) but they also have average Ba-F bond length around 2.55 Å which is also very close to bulk value of Ba-F as 2.66 Å. However, since they are finite size systems, they do not have the translational symmetry. The most intense IR and Raman bands are scattered around 150-190 and 250-310 cm$^{-1}$ frequency region respectively. However, even though $(BaF_2)_{5-6}$ being finite size system, their vibrational behavior are found to be quite similar to what observed for the impurity doped bulk $BaF_2$ crystal with reduced translation symmetry.

## CONCLUSIONS

The calculated Raman and Infrared (IR) spectra of the $(BaF_2)_n$ (n=1-6) clusters are reported. The shift of multiple peaks in Raman and IR spectra towards lower frequency is predicted with the increase in the cluster size. The vibration behavior of these clusters has been compared with the bulk $BaF_2$ ionic crystal. It is found that small size cluster displays behavior closer to the bulk crystal.

## ACKNOWLEDGMENTS

Authors are highly thankful to the department of science and technology (DST), India for funding under HFIBF project. Department of Physics, Michigan Technological University is highly acknowledged for funding for exchange visitor program. Authors are also thankful to Subhash Pingle from Pune University, India, Sanjeev Gupta and Haiying He from Michigan Technological University for valuable suggestions and for help in calculations.

## REFERENCES

[1.] J. Vail, E. Emberly, T. Lu, M. Gu and R. Pandey, *Physical Review B* 57 (2), 764 (1998).
[2.] H. Jiang, R. Pandey, C. Darrigan and M. Rérat, *Journal of Physics: Condensed Matter* 15 (4), 709 (2003).
[3.] R. K. Pandey, M. Kumar, U. B. Singh, S. A. Khan, D. K. Avasthi and A. C. Pandey, *Nuclear Instruments and Methods in Physics Research Section B: Beam Interactions with Materials and Atoms* 314 21–25 (2013).
[4.] M. e. Frisch, G. Trucks, H. B. Schlegel, G. Scuseria, M. Robb, J. Cheeseman, G. Scalmani, V. Barone, B. Mennucci and G. e. Petersson, (Gaussian, Inc. Wallingford, CT, 2009).
[5.] J. Leger, J. Haines, A. Atouf, O. Schulte and S. Hull, *Physical Review B* 52 (18), 13247 (1995).
[6.] M. Mérawa, M. Llunell, R. Orlando, M. Gelize-Duvignau and R. Dovesi, *Chemical Physics Letters* 368 (1–2), 7-11 (2003).
[7.] F. Kadlec, P. Simon and N. Raimboux, *Journal of Physics and Chemistry of Solids* 60 (7), 861-866 (1999).